# Methods of measuring 3D wind-wave spectra based on the sea-surface video image processing

Boris M. Salin and Mikhail B. Salin[1]

*Abstract* - Optical instruments for measuring surface-wave characteristics provide a better spatial and temporal resolution than other methods, but they face difficulties while converting the results of indirect measurements into absolute levels of the waves. We have solved this problem to some extent. In this paper, we propose an optical method for measuring the 3D power spectral density of the surface waves and spatio-temporal samples of the wave profiles. The method involves, first, synchronous recording of the brightness field over a patch of a rough surface and measurement of surface oscillations at one or more points and, second, filtering of the spatial image spectrum. Filter parameters are chosen to maximize the correlation of the surface oscillations recovered and measured at one or two points. In addition to the measurement procedure, the paper provides experimental results of measuring multidimensional spectra of roughness, which generally agree with theoretical expectations and the results of other authors.

## I. Introduction

The problems related to the dynamics of wind-generated waves and their study under field and laboratory conditions remain topical in science. The presence of certain gaps in our knowledge of wave processes motivates the development of a new instrumentation and more sophisticated experimental studies.

---

[1] This work was supported by the Ministry of Education and Science of the Russian Federation (project No. 14.132.21.1428) and the Russian Foundation for Basic Research (project Nos. 11-02-01216a and 13-02-00932a).

The authors are with the Institute of Applied Physics of the Russian Academy of Sciences, 46 Ul'yanov Street, 603950, Nizhny Novgorod, Russia (e-mail: mikesalin@hydro.appl.sci-nnov.ru )

The measurement problem is specific in that nowadays the local characteristics (measured at a single point) of surface roughness are the only parameters that are well studied and clearly understood, and the slope spectrum is slightly less studied. A large amount of experimental data on these parameters has been accumulated, and the corresponding theoretical models have been developed (see, e.g., [1]). A good information base on these parameters is due to the existence of simple local instruments for measurement of wave parameters, which include string wave gauges (individual and grouped), wave rider buoys with acceleration and tilt sensors, vertically oriented sonars, and other tools that provide control of the wave parameters at specific points of the water area [2-5].

As for two-dimensional (frequency - angle), and especially three-dimensional (temporal frequency and two projections of the spatial frequency) characteristics of wind waves, the amount of reliable experimental data on such spectra is significantly less than on one-dimensional spectra. The existing methods and instruments for measuring the angular dependence of the wind-wave spectrum still require further development and improvement[2]. Little data are available on a dispersion relation for the real ocean waves, which is represented as a function of temporal frequency $f$ and spatial frequency $K$. The data on nonlinear characteristics of the wind-wave components are not sufficient.[3]

---

[2] Wave rider buoys or local groups of string wave gauges (which are the most common measurement tools) make it possible to determine only the main or average direction of wind waves. These methods of measurement are based on signal processing techniques that are similar to the maximum likelihood method [3].

[3] Nonlinear effects (and related problems of measurement) include the formation of the group structure of waves [12], modulation of short waves in the presence of intense long waves, the presence of bounded waves, which are due, e.g., to the difference between the form of a propagating wave and a sine curve.

At the same time, information on three-dimensional characteristics of wind waves is needed to solve a number of fundamental problems. For example, detailed information about the three-dimensional spectrum of wind waves is required in underwater acoustics to calculate reverberation parameters for Doppler sonars [7,8]. While studying the friction coefficient, which determines the interaction of the waves and the wind, it is necessary to know whether linear or nonlinear harmonics prevail in the short-wave range [6]. In radar or high-frequency acoustic problems, data and statistics on current wave profiles measured at different wind speeds are needed to determine the scattering signal [5,9,10].

The following brief review is given in order to indicate the place of this paper in the overall context of current research of multi-dimensional characteristics of wind waves. To obtain 3D spectra one needs to continuously and sufficiently frequently (about several Hz) measure 2D profiles of the rough surface. Estimates[4] show that the number of points at which one needs to simultaneously measure the current value of the elevation or slope angle of the surface is about $10^6$. Obviously, none of the contact measuring tools are suitable for this huge amount of points (the maximum amount of string wave gauges reported to work synchronously is 15 to 20).

---

[4] The required number of measurement points $10^6$ is obtained in the following way. Assume that the maximum size of the wave group does not exceed 10 wavelengths. According to this, we choose the size of the spatial analysis window of 10 wavelengths at the lowest temporal frequency. For the analysis of nonlinear phenomena one needs to observe at least several (e.g., five) harmonics of the fundamental temporal frequency in the signal spectrum. If the spatial frequency $K$ is proportional to $\Omega^2$, this means that the spatial coordinate resolution must be at least 1/25 of the length of the low-frequency wave, or 1/250 of the linear size of the aperture. Due to the fact that the minimum number of points is four per wavelength, the required resolution (in points) is 1/1000 of the linear size of the aperture, or ~ $10^6$ points over the whole observation area.

Among the remote sensing facilities that allow direct measurement of the wavy surface oscillations at many points, the LIDAR systems are most suitable (primarily, due to their high spatial resolution) [11]. A limitation of the LIDAR systems, which scan by two angular coordinates, is a relatively long time of observation due to the serial beam scanning. For $10^6$ points and a measurement distance of about 100-150 m, this time is ~1 s, which is not sufficient to monitor the wave processes with frequencies above 0.2 - 0.3 Hz. However, snapshotting a two-dimensional profile of the sea surface and then calculating once the spatial spectrum of wind waves are feasible for the LIDAR systems, including airborne ones [11].

At the same time, relatively simple indirect schemes of obtaining information about the wind waves are widely spread, and they are based on processing of radar (including satellite-borne) and optical images of the sea surface. For example, radar systems are used extensively to estimate an average wave height and other characteristics over large areas [12-14, 20].

Optical indirect schemes of measurement have almost no limitations on the spatial and temporal resolution, because even consumer camcorders record video in "Full HD" mode with ~ 2 megapixels, and with more than 10 megapixels if the mode of a series of frames or other special modes are used. Optical measurement methods can be divided into forward [15] (where the refracted light is recorded), stereoscopic [16,17], and brightness-based. The latter technique is based on a functional dependence between the local brightness of the surface patches and their angle relative to the horizon [18-29].

A detailed discussion of the advantages and disadvantages of all methods, listed above, would require publishing one more paper in the journal. The "brightness" schemes can provide measurement of the wind-wave spectra in the widest range of spatial and temporal frequencies. The development of such techniques has been ongoing for a long time, since [18-22], but their use is still limited by their principle drawback, i.e. the surface brightness is a very complicated function of its form; and it is impossible to set the universal functional, because it greatly varies under different experimental conditions. The brightness formation law is more deterministic (close to the delta function) in the special case of observation of sun glitter (which is not discussed here) [18,23]. Also, in some cases the analysis of the intensities of two light polarizations can give more information [24,25].

Measurements in a wide range of lighting conditions are possible with the help of a method, which is described in this paper and which is based on the classic video shooting (with one lens), that is supplemented by the original algorithm to automatically determine the parameters of the brightness-slope function under the current conditions of imaging. This is accomplished by means of coherent processing of an image series together with a reference record of oscillations in a check point. Data processing algorithm includes spatio-temporal image filtering, a self-test of a result of the optical measurement and error estimation. In our opinion, this algorithm makes it possible to recover the surface roughness parameters more accurately and more fully than in the other papers on the "classic" photo and video shooting of a sea surface [20,26], while our algorithm uses virtually the same input data. The measured values are a three-dimensional spectrum of roughness (as a function of a temporal frequency and two projections of the wave vector), and current wave profiles.

The first results of this project were reported in [28]. This paper was partially presented in the conference proceedings [29] and describes the advanced scheme of video processing and new data on the measurement of 3D characteristics of wind waves.

## II. Functional dependence between an optical image of the sea surface and a spatial structure of the wave profile

Assume that the angular (spatial) dependence of the sky brightness is smooth, the sea surface is observed far from the specular point, and the angles of slope of surface waves are small. Then a linear approximation can be used to derive a simple relation that describes the intensity $U(x,y,t)$ of the optical signal reflected by the sea surface as a function of projections of the slope angles at the point with coordinates $(x, y)$ [21,22]:

$$U(x,y,t) = C_1 \cdot \alpha(x,y,t) + C_2 \cdot \beta(x,y,t), \qquad (1)$$

where $U(x,y,t)$ is the intensity of the video image of a wavy surface (in certain units, without the constant component and sampling variables), which corresponds to the point of the water area with coordinates $(x, y)$, $\alpha$ and $\beta$ describe the time dependence of the projections of slope angles at the point $x, y$ in the longitudinal (from the camera to a scattering area) and transverse directions, respectively, and $C_1$ and $C_2$ are the transformation coefficients.

The coefficient $C_1$ (longitudinal direction) is determined, first, by the angular dependence of the specular reflection from the water surface (Fresnel formulas) and, second, by the brightness gradient in the vertical direction at the specular point in the upper hemisphere, which illuminates the water surface (see Fig. 1). The coefficient $C_2$ (typically, $C_2 \ll C_1$) is determined only by the brightness gradient (in the horizontal direction at the same point of the hemisphere) multiplied by the cosine of the average angle of incidence $\alpha_o$ (Fig. 1). In the case of large $\alpha_o$, the waves coming from the transverse direction will hardly be noticeable on the brightness image of the surface. In this case, the brightness of the image can be directly converted into a spatial distribution of angle projections in the longitudinal direction.

Upon replacement of small angles α and β by sinα and sinβ, Eq. (1) can be written as a dot product of the gradient of the current altitude of surface roughness $H(x, y, t)$ and a dimensional vector **C**, which has projections $C_1$ and $C_2$ on the longitudinal and transverse axes, respectively:

$$U(x,y,t) = (\nabla H(x,y,t), \mathbf{C}). \qquad (2)$$

Of course, the surface brightness is a very complicated function of its form, and this function if much more complicated than (1,2). In fact, we make an assumption that a series of distorting effects can be neglected and the brightness-slope function can be considered as linear. An image processing algorithm will be built below on the basis of this hypothesis. The algorithm includes a self-test. The hypothesis of linearity will be checked each time, basing on the results of this self-test.

### III. Measurement scheme

The proposed measurement scheme requires at least two synchronous video records of the sea surface. First, a patch of a rough surface over an area of 10x10 or more wavelengths of the energy carrier waves is imaged (see Fig. 2a). This provides a 3D (two coordinates and time) set of data on the surface brightness. Second, in order to measure the local surface oscillations (similar to measurements by a string wave gauge), one makes a zoomed video record of one or more so-called optical spar buoys [28] located in the same area (see Fig. 2c). These local measurements are needed to calibrate a video image when it is transformed into a time function of the wave profile.

According to the theoretical assumptions the proposed method works under conditions of a uniform cloud cover, and at a clear sky. The video images of the surface should be done outside the solar glare and preferably with no foam crests of the waves.[5]

The method of measurement uses off-shelf "semi-professional" camcorders. The following ones were utilized in the experiments described below: Sony DCR-DVD405 with 1024x576 frame size and recording to a mini-DVD, as well as JVC Everio GZ-HM200 with 1900x1600 frame size and writing to a flash card.

The method of measuring surface oscillations with optical spar buoys is described in sufficient detail in [28] and is not described here. The spar buoys are equipped with an anchor, buoyancy, and standards of length in the vertical and, optionally, horizontal directions (in the latter case, it is a disk of known diameter). In the case of deep water, the spar buoys can be made free floating, with a load to maintain the vertical position and a dynamic damper, tuned to a certain frequency limit. In principle, the landmarks can be replaced by other small-sized sensors if they are calibrated, and they are synchronized with the video.

Video image of a large surface area (Fig. 2a) is processed in the following way. First, an image is subjected to geometric transformation by scaling (conversion of pixels into meters) and correction of geometric distortion (perspective). Camera shake is also compensated. Two transverse and one longitudinal reference sizes in the frame are enough for correct scaling. When working in stationary conditions (video recording from bridges, coastal structures, etc.), the scaling can be done in advance by moving one time a reference over the water area.

---

[5] By this moment, the method does not take into account the foam caps on the waves, which break a specular reflection of light. The proposed algorithm needs to be modified to consider the foam caps.

These operations result in a surface brightness field $J(x, y, t)$ expressed in arbitrary units (for example, in volts) and tied to the area coordinates $x, y$ and time $t$.

Next, the image is subjected to Fourier transform in three dimensions (two spatial coordinates and time) for each time interval of $T \sim 20$ s with overlapping (the duration is determined by the needed resolution of the wind-wave spectrum):

$$I(t_n, \mathbf{K}, \Omega) = \frac{1}{L_x L_y T} \int_{t_n}^{t_n+T} \int_0^{L_x} \int_0^{L_y} J(x, y, t) \exp(-iK_x x - iK_y y + i\Omega(t - t_n)) \\ \times w_x(x) w_y(y) w_t(t - t_n) dt \cdot dx \cdot dy, \qquad (3)$$

where $I(t_n, \mathbf{K}, \Omega)$ is the current value of a 3D spectrum of brightness as a function of the coordinates, namely, the wave vector $\mathbf{K} = \{K_x, K_y\}$ and the frequency $f = \Omega/(2\pi)$, $L_x$ and $L_y$ are the window sizes in $x$ and $y$, respectively, $T$ is the duration of a time window, and $w_x$, $w_y$, and $w_t$ are window functions. Hann, sinusoidal, or rectangular windows will be used below.

Spatial spectra are used due to the simplicity of transformation from the gradient to elevation in the spectral domain and then to the desired function, i.e., current values of the wave profile $H(x,y,t)$, using the inverse Fourier transform.

The current values of 3D spectra of elevation (wind-wave spectra), which are defined as $G(t_n, \mathbf{K}, \Omega) = \mathbb{F}[H(x, y, t)]$, where $\mathbb{F}$ is the Fourier transform given by Eq. (3) and the time $t$ lies in the range $t_n < t < t_n + T$, will be sought for positive[6] $\Omega$ as the product of a brightness spectrum and a transform coefficient $\Phi(\mathbf{K}, \Omega)$:

$$G(t_n, \mathbf{K}, \Omega) = \Phi(\mathbf{K}, \Omega) \cdot I(t_n, \mathbf{K}, \Omega). \qquad (4)$$

---

[6] $G(t_n, \mathbf{K}, -\Omega) = G^*(t_n, -\mathbf{K}, \Omega)$ is the Fourier transform of a real-valued function. If the sign in the exponent is chosen as in Eq. (3), then the distribution of $G^2$ will be correctly oriented along the $K_x$ and $K_y$ axes at $\Omega > 0$, in the same way as the waves are oriented along the $x$ and $y$ axes on the video.

The way of determining Φ(**K**,Ω) is discussed in Sec. 4. The function Φ(**K**,Ω) (hereafter called a calibration function) can be verified, for example, by the maximum of the correlation coefficient (see Sec. 4) of the recovered elevation field and the reference signal, which was obtained with an optical spar buoy.

The 3D power spectral density (PSD) of wind waves $G^2(\mathbf{K},\Omega)$ will be calculated in a standard way by incoherent accumulation of the current values of 3D wind-wave spectra over time (~ 20 samples (4)):

$$G^2(\mathbf{K},\Omega) = \frac{TL_xL_y}{2\pi^2} \cdot \frac{1}{N}\sum_{n=1}^{N}|G(t_n,\mathbf{K},\Omega)|^2 . \tag{5}$$

Hereafter the symbol $G^2$ ($I^2$) denotes PSD, a quadratic value, which is normalized so that $\sigma^2 = \int_0^\infty df \int_{-\infty}^\infty dK_x \int_{-\infty}^\infty dK_y \cdot G^2(\mathbf{K},2\pi f)$, where σ is a standard deviation of the surface. The coefficients in Eq. (5) have such a value due to the fact that one channel of the 3D spectral analysis has a resolution $\Delta K_{x,y}=2\pi/L_{x,y}$ by the wave vector and $\Delta f=1/T$ by the frequency, and only the positive frequencies are considered in the normalization condition. If the goal of the image processing is to calculate PSD, then the Hann function is used as $w_t$ in Eq. (3), and the sinusoidal function is used as $w_x$ and $w_y$.

If the goal of the image processing is to recover the current wave profiles, then no special window functions are used in Eq. (3). After calculating (4) we compute the inverse Fourier transform $H(x,y,t) = \mathbb{F}^{-1}\left[G(t_n,\mathbf{K},\Omega)\right]$.

One can note that the most complete description of statistical characteristics of wind waves is achieved by computing the 3D PSD $G^2(\mathbf{K},\Omega)$. In a number of studies (see, e.g., [30]), the results of measurements and calculations are presented in the form of a 2D PSD as a function of frequency and angle of $G^2(\theta,\Omega)$, while the modulus of the wave vector is given by the dispersion relation at each frequency. Nonlinear wave components are not described in this case, so the authors of this paper do not use this form of data representation.

Processing of a video file by the above scheme takes a factor of 1.5 to 2 longer time than the duration of this file.[7] Therefore, the developed measurement scheme can be used to obtain rapid information during in-situ experiments. Other special methods of image processing can also be employed in addition to the described one to improve the quality of the output [31].

### IV. Methods of converting a brightness field into an elevation field using the calibration function

Equation (2) can be rewritten in the following form:

$$J = q(\nabla H, \mathbf{s}) + J_0, \qquad (6)$$

where $J_0$ is a constant background level, which is independent of the position and time, $q$ is a dimensional amplitude factor, and $\mathbf{s}$ is a dimensionless unit vector in the horizontal plane, whose direction is close to the azimuthal direction of the camera sight. Thus, in fact, the equation which relates oscillations and brightness contains only two unknown parameters, which are the normalizing factor $q$ and the direction of the vector $\mathbf{s}$.

---

[7] Comments on the current program. The described approach has been implemented in a C++ program with the use of OpenMP for running multiprocessor (multicore) computers. Camcorders record data on conventional media such as DVD drives or flash memory cards in a standard video format. These video files are decomposed into frames using a free program Ffmpeg, i.e., converted into a folder with image files of a standard format. The image files can already be read by any standard library, e.g., the free library wxWidgets. The free library FFTW is used to calculate the Fourier transform. Links to the mentioned libraries can easily be found on the Internet.

If the direction of the vector **s** is known, and the axis $x$ is chosen along it, then according to Eq. (6), one can simply integrate the brightness field by Eq. (7) in order to find the profiles of the waves (elevation field) $H(x,y,t)$ with accuracy up to a constant factor $q$:

$$H(x,y,t) - \frac{1}{L}\int_0^L H(x',y,t)dx' = \frac{1}{q}\int_0^x (J(x',y,t) - J_0)dx' - \frac{1}{qL}\int_0^L\int_0^{x'} (J(x'',y,t) - J_0)dx'dx''. \quad (7)$$

Here, $\frac{1}{L}\int_0^L H(x',y,t)dx'$ is an average by $x$ value of the wave profile at the time moment $t$.

Equation (7) can be simplified when the following condition is satisfied:

$$\overline{|H(x,y,t)|^2} \gg \overline{\left|\frac{1}{L}\int_0^L H(x',y,t)dx'\right|^2}, \quad (8)$$

which means that a rather large amount of wavelengths (or their projections) can be fitted along the $x$ axis of the measurement aperture of length $L$. In this case, Eq. (7) takes the form containing an integral of the brightness field, which is known:

$$H(x,y,t) = \frac{1}{q}\int_0^x (J(x',y,t) - J_0)dx' - \frac{1}{qL}\int_0^L\int_0^{x'} (J(x'',y,t) - J_0)dx'dx''. \quad (7^*)$$

Condition (8) is satisfied in the case of the correct choice of the observation angle, where the axis of the camera coincides with the main direction of propagation of the waves. However, inequality (8) is not satisfied, and expression (7*) is invalid if the energy of the waves is focused in the direction orthogonal to the axis of the camera sight. In this case, full recovery of the wave profile is apparently possible only when the video is shot from two points of view, whose angles are orthogonal, and the obtained wave profiles are accumulated in a coherent way.

In the case of a single point of view, the accuracy of recovery of the wave profiles can be increased by turning to processing in the spectral domain. The amplitudes of the spatial harmonics should be corrected by a function of the form $1/\cos\theta$ (here, θ is an azimuthal angle relative to the vector **s**).[8] Then the wind waves, which come from the side directions and are almost invisible in the video, will be strengthened. To carry out this angular correction, a 3D spectral analysis (3) is used, followed by optimal filtering [31] in the wave number and frequency domains.

**Energy calibration**. At first, for comparison, we consider a simplified so-called "energy" approach to calibration (transformation) of the video, which can be used in the case of absence of a record of surface oscillations at a point that is synchronous with the image.[9]

Within this measurement scheme, one should look for the calibration function in the form of the following function of the modulus **K**:

$$\Phi(\mathbf{K},\Omega) \approx A/K. \tag{9}$$

(here, both $K$ and $\Omega$ are considered to lie within the operating range). Using a transfer function in the form $\sim 1/K$ (similar to integration), we convert the amplitudes of the slope angles into elevation amplitudes. Constant $A$, which is an unknown parameter, can be found by minimizing the expression[10]

$$\int_{\Omega_1}^{\Omega_2} \left| G^2(\Omega) - G_{buoy}^2(\Omega) \right| d\Omega \to \min, \tag{10}$$

where $G^2(\Omega) = \iint G^2(\mathbf{K},\Omega) dK_x dK_y = \iint \Phi^2(\mathbf{K},\Omega) I^2(\mathbf{K},\Omega) dK_x dK_y$ is the calculated average wind-wave PSD in the water area in which the image is recorded, $G_{buoy}^2(\Omega)$ is the wind-wave

---

[8] Generally speaking, we have not exactly compared the precision of integral (7) with the spectral approach.

[9] Due to the fact that no precise self-examination is available without the correlation processing, such a wave characteristics recovery scheme, obviously, will be approximate and will not consider the direction of the vector **s**.

[10] Generally speaking, $G^2_{buoy}(\Omega)$ in Eq. (10) should be preceded by a factor, whose value is $0<\eta<1$, because according to Eq. (2), the video is formed by only one projection of the wave slope angle, and the recovered spectrum $G^2(\Omega)$ is only a part of the total spectrum $G^2_{buoy}(\Omega)$.

PSD measured at a single point, and $\Omega_1..\Omega_2$ is the frequency domain which contains most of the wind-wave energy (or the region of interest).

Representation of the calibration function in the form (9), independent of the frequency $\Omega$, follows from the physical sense, according to which the amplitude of the specular reflected light signal depends only on the incidence-reflection angle and is independent of the rate (frequency) of its variation.

When this approach is applied to the actual conditions, in order to avoid singularity at low spatial frequencies ($K \to 0$) and to reduce the harmonic effects, the transfer function is sought in the following modified form:

$$\Phi(\mathbf{K},\Omega) = \Phi_K(K,\Omega) = \frac{A \cdot K}{K^2 + \max\left(\left(\gamma \cdot \Omega^2 / g\right)^2;\ K_{min}^2\right)}, \tag{11}$$

where $g$ is the free-fall acceleration. In this case, Eq. (10) determines not only the constant $A$, but also other variable parameters, which are $\gamma$ and $K_{min}$. Experiments show that one can preselect $\gamma=1/4$, $K_{min}=2\cdot\Delta K$, $\Omega_1\approx\Omega_p/2$, and $\Omega_2\approx5\Omega_p$ as typical parameters (here, $\Omega_p$ is the angular frequency of the wind-wave spectrum maximum and $\Delta K$ is the resolution of the spectrum in $K$).

The function $\Phi_K(K,\Omega)$ retains the qualitative features of Eq. (9), i.e., it decays as $\sim 1/K$ in the short-wave region, but $\Phi_K$ is saturated in the long-wave area. Practice has shown that the saturation should be frequency dependent. The cutoff region lies at a constant level $\gamma$ relative to the wave number defined by the dispersion relation

$$K_{theor} = \Omega^2 / g. \tag{12}$$

One should include an additional frequency adjustment $W(\Omega)$ for very low frequencies

$$\Phi(\mathbf{K},\Omega) = W(\Omega)\Phi_K(K,\Omega) \text{ if } \Omega<\Omega_1. \tag{13}$$

This strictly enforces the condition $G^2(\Omega)=G^2_{buoy}(\Omega)$ if $\Omega<\Omega_1$.

**Use of synchronous recording.** If the record of wind-wave images is synchronized with a reference record of local surface oscillations at a point within the field of view, then the processing scheme may be changed.

In this case, the calibration function, which already depends not only on the magnitude, but also on the direction of the vector **K**, should be sought, as in Eq. (11), in the form

$$\Phi(\mathbf{K},\Omega) = A \cdot \frac{-i(\mathbf{K},\mathbf{s})}{(\mathbf{K},\mathbf{s})^2 + (K\sin\delta)^2 + K_{min}^2}, \qquad (14)$$

where $K_{min}$ and $\delta$ are some parameters built into the calibration function for its smoothing and limitation. Namely, the parameter $\delta$ eliminates the singularity in the direction orthogonal to **s**, and the parameter $K_{min}$, as in Eq. (11), cuts off a long-wave part of the spectrum, which is described by small waves on a characteristic size of the region and has a low signal-to-noise ratio. For $|\mathbf{K},\mathbf{s}|>>K_{min}$, function (14) tends to its optimal form

$$\Phi(\mathbf{K},\Omega) \approx \frac{-i \cdot A}{(\mathbf{K},\mathbf{s})}, \qquad (15)$$

which follows from Eqs. (6) and (7*).

The newly introduced parameters $A$, **s**, $K_{min}$, and $\delta$ can be found from two conditions. The first condition is Eq. (10), and the second one is maximization of the modulus of the correlation coefficient $B(\Omega)$. The latter is defined by Eq. (16) in the form of a correlation between the displacement signal $H_{buoy}(t_n,\Omega)$, which is directly measured at the point $(x', y')$, and the displacement signal $H_{video}(t_n,x',y',\Omega)$, which is recovered at the same point as a result of processing of the video:

$$B(\Omega) = \frac{\sum_n H^*_{buoy}(t_n, \Omega) H_{video}(t_n, x', y', \Omega)}{\sqrt{\sum_n \left|H_{buoy}(t_n, \Omega)\right|^2 \cdot \sum_n \left|H_{video}(t_n, x', y', \Omega)\right|^2}}, \qquad (16)$$

where $(x', y')$ are the coordinates of the "optical" spar buoy; $H_{buoy}(t_n, \Omega) = \mathbb{F}_t\left[H_{buoy}(t)\right]$ are the current values of a complex spectrum of the signal, which is measured by the spar buoy; $H_{video}(t_n, x', y', \Omega) = \mathbb{F}_\mathbf{K}^{-1}\left[G(t_n, \mathbf{K}, \Omega)\right]$ is the result of processing of the brightness image by Eqs. (3) and (4), followed by the inverse Fourier transform of $G(t_n, \mathbf{K}, \Omega)$ in $\mathbf{K}$ and (finally) by selecting the values of the obtained function at the point of an "optical" spar buoy.

The summation in Eq. (16) is carried out over the number $n$ of time windows. In order to calculate $B(\Omega)$, the time window size should be selected such that the total number of windows (maximum number of terms in Eq. (16)) is no less than 30-50. Thus, the frequency resolution is usually ~ 0.1-0.2 Hz when the correlation function is sought. The maximum $|B(\Omega)|$ in $\mathbf{s}$ and $K_{min}$ is found by simple sorting. Amplitude factor $A$ is determined mainly[11] by energy equation (10).

A weak frequency function of the parameters $A(\Omega)$, $\mathbf{s}(\Omega)$, and $K_{min}(\Omega)$ can be introduced for increasing the correlation coefficient and for a better agreement of the recovered function $H_{video}(t_n, x', y', \Omega)$ with the direct experimental data. In this case, it is important to have a long record of the signal to increase the number of independent samples.

---

[11] In the case of a spatially inhomogeneous profile of wind waves, it is possible to cover the area where the average value of $G^2(\Omega)$ is determined.

## V. Examples of measurements and some experimental results

The described method of measuring wave characteristics has been tested in a series of experiments in the Baltic Sea and Lake Ladoga. The experiments were performed under stationary conditions, near the coast, in closed bays; thus, low levels of wind waves were generally observed. The parameters of the obtained video records, which will be discussed below, are presented in Table 1.

Obviously, $\lambda_{min}$ is not a fixed parameter, and it is determined by the distance to the camera and its zoom. One can speak of a fixed ratio of the lengths of the maximum and minimum measured waves. This ratio is defined by the resolution of the video recorder and by the desired number of the maximum wavelengths in the frame (i.e., selected zoom). In these experiments, the shooting was carried out from a relatively low height and with a low angle to the horizon; thus, the values of $\lambda_{min}$ were different on the $x$ and $y$ axes (Table 1 indicates the lowest value).

**Energy approach.** First, this section presents the experimental results on three-dimensional spectra of wind waves, which were obtained using the "energy" calibration and transformation given by Eqs.(10) and (11). According to the processing scheme, the video record is converted into a 3D spectrum of wind waves by exploiting spatial spectral analysis of video and then minimizing differences between two samples of one-dimensional wind-wave PSD. The first sample is measured directly. The second one is recovered on the basis of the brightness field.

Figure 3 shows these one-dimensional wind-wave PSDs. It is seen in this figure that good agreement between measured and recovered frequency functions can be achieved by varying the values of a small number of parameters ($A$, $K_{min}$, and $\gamma$). Mismatch between the maxima of the curves in Fig. 4a (experiment I) is due apparently to the fact that the measurements were done in different time periods. In this case, video was calibrated by comparing high-frequency "tails," whose dependence on the current value of the wind speed is usually weaker.

Generally speaking, small differences can and should occur, because the recovered spectra were obtained by averaging over the entire space, while the spar buoy data were obtained at a local point.

Frequency $\Omega_1$ is about one-half of the signal peak. For $\Omega<\Omega_1$, frequency correction (13) has to be applied; therefore, the curves completely merge below frequency $\Omega_1$. It should be noted that this area is little informative in terms of the study of wave processes.

Theoretical functions describing the Toba spectrum (17) and high-frequency asymptotic forms of the Pearson-Moskowitz spectrum (18) were also plotted for comparison [1]. These functions usually correspond to the frequency range above $1.5\Omega_p$. According to the plots, the measurement results are in better agreement with the Toba spectrum:

$$G_{Toba}^2(\Omega) = 2\pi\alpha_T g u_* \Omega^{-4} \tag{17}$$

$$G_P^2(\Omega) = 2\pi\alpha_p g^2 \Omega^{-5}, \tag{18}$$

where $\alpha_T = 0.127$ is the Toba constant, $\alpha_p = 8.1 \cdot 10^{-3}$ is the Philips constant, and $u_*$ is the friction velocity of wind. The factor $2\pi$ was included in both cases to make the corresponding values normalized to $\Delta f$. Thus, the quantities given by Eqs. (17) and (18) are measured in m$^2$/Hz.

Unfortunately, the wind speed was not measured in the experiment, and the values of $u_*$, which were used for the construction of theoretical functions, were selected according to the experimental plots of the wind-wave PSD. For a reference, the wind speed ($U_{10}$) at a 10 m height is provided by converting from $u_*$ by the formula

$$U_{10} = \frac{u_*}{\kappa} \ln \frac{10\text{m} \cdot g}{\alpha_{CH} u_*^2}, \tag{19}$$

where $\kappa=0.4$ is the von Karman constant and $\alpha_{CH}=0.0144$ is the Charnock constant.

Figure 4 shows a three-dimensional function of the PSD in the form of the spectrum cross sections as functions of $K_x$ and $K_y$ at certain frequencies. Hereafter, the $x$ axis is the horizontal axis in the frame, and the $y$ axis is the camera viewing direction. According to the plots, the surface waves come from different directions (possibly due to the reflection from the closely located coast). For record I (Figs. 4a-4c), the width of the main lobe was estimated by standard interpolation of an angular function of the form $(1+\cos\theta)^n$, and $n$ was about 20.

Analysis of the figures shows that the maxima of the spatial spectra at a fixed temporal frequency are mainly located near the rings defined by dispersion relation (12). Analytical estimation of the width of the ring shown in Fig. 4 is based on the resolution of the spectral analysis either in **K** or $\Omega$. In the second case, the wave vectors $K$ are chosen on the basis of dispersion relation (12) at each frequency band $\Omega\pm\Delta\Omega/2$. In these plots, the borders of $K$ defined due to $\Delta\Omega$ turned out to be larger than the spatial resolution of the spectral analysis.

Record III (Figs. 4g-4i) is an example of the distortion of the dispersion relation of surface wind waves due to the presence of wind-induced or other currents. The experimentally measured spectra in Figs. 4g-4i are compared with the dispersion relation plots for a still water (12) and in the presence of a current.

In the presence of current at a rate $w$, Eq. (12) is converted to an equation that determines all possible wave vectors **K** at each fixed frequency $\Omega$:

$$w^2 K^2 \cos^2\varphi - (2w\Omega\cos\varphi + g)K + \Omega^2 = 0. \tag{20}$$

Here, the modulus $K$ and the angle $\varphi$ are the polar coordinates of **K**, and the angle $\varphi$ is taken relative to the flow direction. Equation (20) passes into the simpler form (21) if $w$ is small ($g/\Omega \gg w$), i.e., when the phase velocity of surface wind waves is higher than the speed of current:

$$K = \frac{\Omega^2}{g}\left(1 - 2\frac{\Omega}{g}w\cos\varphi\right). \tag{21}$$

Dispersion curves (20) and (21) are plotted in Figs. 4g-4i. Since the velocity modulus and direction of the current are unknown, they were adjusted for the best match of the theoretical curves and experimental two-dimensional spectra. The obtained value of the velocity modulus is 0.38 m/s. A qualitative effect is observed both in theory and in the experiment. The curves on the $K_x$, $K_y$ plane are no longer closed starting from a critical frequency.

One can observe nonzero levels of the spectra in Fig. 4 at points whose coordinates do not satisfy the dispersion relation. This may be due to several reasons, including the presence of non-linear processes, both in surface roughness and in transformation of the wave slopes into brightness.

The spectrum can be averaged in an angular sector and plotted as a function of the wave number modulus $K$ and frequency $\Omega/(2\pi)$ in order to study the nonlinear phenomena. Such diagrams are provided for record I in Fig. 5, where panel *b* shows the wind-wave PSD and panel *a* shows directly the brightness PSD (in order to isolate the calibration effects).

Energy distribution as a function of the phase velocity of surface waves ($v=\Omega/K$) for a fixed wave number can easily be calculated on the basis of 2D cross sections in the form presented in Fig. 5. Such a function is needed, for example, to estimate the Doppler spectrum of the Bragg backscattering of electromagnetic [10] or acoustic [7,8] waves by wind waves. Such an effect takes place when the horizontal projection of the electromagnetic (acoustic) wave vector **k**$_{xy}$ (which is usually given) satisfies the relation $2\mathbf{k}_{xy} = \mathbf{K}$. Figure 6 shows the discussed phase velocity spectrum for two angular sectors and four values of the wave number modulus (for record II, the respective incident wavelength projection is from 30 cm to 2.5 m). It can be seen that for each $K$ the waves move not only with the velocity $v = \sqrt{g/K}$ determined by the dispersion relation for surface gravity waves, but also there is a set of waves moving with higher or lower velocities.

The observed deviations of the measured PSD distribution from the dispersion relation can be described in a phenomenological way without discussing the nature of the phenomena and divided into the following classes:

- finite bandwidth around the dispersion curve L, which exceeds the resolution of the spectral analysis;
- nonzero background level;
- local maximum of intensity around the curve Q (second harmonic), which is plotted according to the function $\Omega_{2nd}^2 = 2K_{2nd}g$ (this expression is derived from the basic dispersion relation (12), so that the phase velocities of the first and second harmonics coincide);
- slow-wave area A, which can be described as a result of nonlinear subtraction of the main energy-carrying wave with frequency $\Omega_p$ and higher-frequency waves that satisfy the dispersion relation. Curve S is plotted in line with Eq. (A1) derived in Appendix A.
- fast-wave area B, which can be described as a result of nonlinear summation of the main energy-carrying waves with different directions. Since the calibration function contains the energy dependence $K^{-2}$, this local maximum is better seen after the calibration (Fig. 5b) than in the initial spectrum of brightness (Fig. 5a).

In principle, these effects may be a reflection of the actual physical processes in the wind-wave field, but they also may be an error of measurement. Thus, the surface waves of finite amplitude are nonlinear; moreover, the high-frequency waves are permanently in a state close to collapse. However, the gradient-brightness dependence, which is the basis of measurement, is also a nonlinear function. At present, the authors have analyzed and estimated all factors and are inclined to believe that these effects are of a physical nature, except, perhaps, for the local maximum in area B.

The relative levels of nonlinear components in the brightness spectrum (Fig. 5a) are higher than the estimate of the distortion introduced by the nonlinear gradient-brightness dependence in the Fresnel formula (see Appendix B, Fig. B1).

It is noteworthy that other researchers have obtained similar 3D wind-wave spectra that contain qualitatively the same nonlinear components, using radar measurements [12,13] and numerical simulations [32]. Unfortunately, in the case of measurement with a system of closely located string wave gauges, the maximum likelihood method does not ensure that two waves with the same frequency and different (both in direction and in magnitude) wave vectors are recorded in a single time window. Therefore, only one branch in the $K, \Omega$ plane can be seen in the spectra [2,4] measured by string wave gauges. However, due to the observed deviations of this branch from the dispersion relation, one can assume the existence of such nonlinear components in the wind waves. Radars (and optical) facilities can introduce nonlinear distortions in the measurement results, but the numerical model and the wave recorder are free of these faults.

Final conclusions on the nonlinear components of the wind-wave spectrum will be made as a result of further research, which will include the improvement of the experiment accuracy and analysis of the phase relations between the fundamental and higher harmonics.

**Correlation approach**. Experiment II gave an opportunity to make a synchronous video record of the wind-wave surface and a record of the waveforms (water-level variation) at the location of the optical spar buoy. This allowed us to use the calibration function in form (14) or

(15) and find the direction of the unit vector **s**, which maximizes the correlation function (16) of the measured $H_{buoy}(t_n, \Omega)$ and calculated $H_{video}(t_n, x', y', \Omega)$ current values of the complex wind-wave spectrum. Since one of the unknown parameters, direction of the vector **s**, is selected according to the maximum correlation at each frequency $\Omega$, it is expedient to calculate the parameter $A(\Omega)$ at all frequencies too, using the expression $\iint \Phi^2(\mathbf{K}, \Omega) I^2(\mathbf{K}, \Omega) dK_x dK_y = G^2_{buoy}(\Omega)$, which is similar to Eq. (10).

Figure 7 shows the modulus of correlation coefficient (16) as a function of frequency, which was obtained for different types of calibration function.[12] The solid black line shows the correlation coefficient for the optimal type of calibration function[13] (14). The correlation coefficients range from 0.5 to 0.8. The solid gray line shows the correlation coefficients for a more simplified form of the gauge function: $\Phi(\mathbf{K}, \Omega) \sim i \cdot A \cdot sign(\mathbf{K}, \mathbf{s})/K$. In this case, there is no angular dependence of the modulus of the calibration function, including the absence of a singularity for the two directions orthogonal to **s**.

It should be mentioned that the correlation coefficient does not decrease significantly if only the phase (sign) of $\Phi(\mathbf{K}, \Omega)$ is angle dependent. At the same time, the correlation function is greatly reduced to the level of 0.2 - 0.4 in the case of the total absence of the angular dependence of the calibration function. The correlation between the measured surface oscillations and the image itself (brightness of a single pixel) is approximately at the same low level (see two bottom lines in Fig. 7a).

---

[12] Only a typical dependence of the calibration function is indicated in Fig. 7a to make the inscriptions shorter. This function was calculated using the formulas given in Sec. 4 containing both saturation and a decay of the function to 0 at the singular points.

[13] The value $\delta=6°$ is used. Hereafter the parameter $K_{min}$ is varied to maximize the correlation coefficient.

Figure 7b presents the results of studying the effectiveness of optimization of the vector **s** direction at each frequency individually. According to this figure, a fixed (average for all frequencies) direction of the vector **s** reduces the correlation coefficient by ~ 0.1 – 0.4.

Figure 8 shows the correlation coefficient as a function of the vector **s** direction for three selected frequencies. It is seen in this figure that the angular dependence has the form of the modulus of sinα with a very irregular main lobe. The main directions of the beam (~ 70º and 250º) are close to the direct and opposite viewing directions of the camera (90º and 270º), and the minima are near 160 and 340º. This generally confirms the correctness of the model which was adopted and used to convert video into the height field.

Considering our studies in chronological order, we note that, initially, we tested the possibility of using a trivial relation between the local brightness and the elevation without applying spatial filtering of images. The correlation coefficient, which is obtained for record II in this case, is 0.25 on the average, and is shown in Fig. 7a by a green solid line (gray dashed line in the black and white version). It should be noted that during the experiments in laboratory tanks (channels), where the waves have a single direction, a "trivial" complex coefficient of the "brightness - elevation" correlation could reach about 0.9 in absolute value with ~±π/2 phase shift. The *in-situ* measured wind-wave spectrum has a large angular width (see Fig. 4d), and the line in which the elevation-brightness function changes phase (and the "blind" zone) divide the spectrum into two parts.

Therefore, the modern wind-wave image processing scheme includes the stage of transition to the spatial spectra and the use of special filters, whose shapes are chosen based on the model of the brightness-into-elevation transformation, and the parameters are chosen (found) by the maximum of the correlation coefficient of recovered wave profiles and initial experimental data on the surface oscillations.

## VI. Future work

The possibility to validate the results of the optical image processing using correlation analysis was demonstrated above. This gives grounds to eliminate distortions introduced by nonlinear dependence slope-brightness. In the future we plan to increase the precision of the experiment and carry out a finer analysis, which will determine the degree of correlation between the second harmonic in the image (Q in Fig. 5) and the reference surface oscillations. In the simplest case, one can decide whether to suppress the second harmonic by a filter or not. In general, the surface brightness model (2) should be supplemented by the quadratic terms with undetermined coefficients. The values of such coefficients can be found in a similar way, and thus a nonlinear calibration function can be built.

Furthermore, in some cases, the phase relation between the fundamental and higher harmonics can help to distinguish nonlinear surface wave from nonlinear optical distortions.

From the technical point of view one can point out the possible ways of development of the proposed measurement scheme. Firstly, in order to increase the surveillance area and record the waves with long periods, a camcorder is going to be installed on the radio-controlled air platform (for example, quadrocopter), using image stabilization by fixed objects in the frame, for example, by the same spar buoys. Secondly, we consider the possibility to arrange joint work of a camcoder and a lidar system with combined optical axes, which would permit fully noncontact measurements. In this case, however, the complex requires reliable stabilization of the platform or stationary setting (on spans of bridges, lighthouses, and shore facilities).

**Conclusions**

In this paper, we present a method for measuring the spatio-temporal characteristics of surface wind waves, namely, averaged 3D spectra and current profiles of the waves. The method records the brightness field of a rough surface using a camcoder and by synchronous (with video) measurement of the wave surface oscillations at one or several locations. The paper gives examples of measurements of multidimensional characteristics of wind waves, which generally conform to the theoretical expectations and the results of other authors.

Some of the effects are associated with the nonlinear spectral components and require further research. Such research becomes possible due to the proposed method of the quantitative estimation of the error levels.

**Appendix A**

Basing on the following arguments, it was possible to derive an analytical function, which defines curve S in Fig. 5 - the boundary of area A, which contains slow nonlinear harmonics. If the superposition of two plane waves with frequencies and wave vectors $\{\Omega_1, \mathbf{K}_1\}$ and $\{\Omega_2, \mathbf{K}_2\}$ (let $\Omega_2 > \Omega_1$) is transformed with a quadratic nonlinearity, then the resulting signal contains different harmonics of the original signal, including the difference wave with the parameters $\Omega_{sub} = \Omega_2 - \Omega_1$ and $\mathbf{K}_{sub} = \mathbf{K}_2 - \mathbf{K}_1$. The nonlinearity can be contained both in the measuring channel and in the wave propagation medium.

The largest value of the nonlinear response is observed when a wave with the lowest frequency belongs to the spectral peak: $\Omega_1 = \Omega_p$. The minimum length of the difference wave vector $K_{sub,min}$ is achieved at collinear $\mathbf{K}_1$ and $\mathbf{K}_2$. For comparison with the observed spectra, $K_{sub,min}$ should be found in the form of a function of the difference frequency $\Omega_{sub}$. Since the initial waves $\{\Omega_1, \mathbf{K}_1\}$ and $\{\Omega_2, \mathbf{K}_2\}$ satisfy dispersion relation (12), the following expression can be obtained:

$$K_{sub}(\Omega_{sub}) = \left(2\Omega_p \Omega_{sub} + \Omega_{sub}^2\right)/g. \tag{A1}$$

Curve S in Fig. 5 was plotted by this formula, and it constrained quite accurately area A. One can note that the curve (A1) starts from zero at an angle $2\Omega_p/g$, which corresponds to the group velocity at a frequency $\Omega_p$. A similar line was observed in [13] in the spectrum measured by the radar, and it is called a "group line" there.

**Appendix B**

Let us check the possibility of using linear approximation while converting the brightness of wind waves images into a field of gradients, based on the following assumptions: a) surface wind waves are set by the standard models, and b) change in the brightness as a function of the local angle of a rough surface slope is calculated by the Fresnel formula [22], according to which the reflection coefficient of unpolarized light is

$$R_{Fr}(\theta_i) = \frac{1}{2}\left(\frac{\sin^2(\theta_i - \theta_t)}{\sin^2(\theta_i + \theta_t)} + \frac{\tan^2(\theta_i - \theta_t)}{\tan^2(\theta_i + \theta_t)}\right), \tag{B1}$$

where $\theta_i$ is the incidence angle reckoned from the normal; $\sin\theta_i / \sin\theta_t = 1.33$.

Consider the optical observation of a rough surface with standard deviation of the slope angles $\sigma$ and at an average angle of incidence $\alpha_o$ of the light beam (see Fig. 1). The ratio of amplitudes[14] of the reflected signals, which are quadratic and linear functions of the slope, can be estimated by the numerical calculation of the spectrum of the function:

---

[14] Here, the amplitude refers to the amplitude at the output of the photo detector. For a light wave, it is intensity.

$$y(t) = R_{Fr}\left(\alpha_o + \sqrt{2}\sigma \sin(\Omega t)\right) \tag{B2}$$

and then by calculation of the amplitude ratio at frequencies $\Omega$ and $2\Omega$. Equation (B2) is the amplitude of the signal reflected from a test wave with frequency $\Omega$ if the incident-wave amplitudes are constant (and equal to 1) in the proper sector of incident angles.

Let $\sigma$ be found based on a model wind-wave spectrum for the sea state corresponding to record I. Since the data on $\sigma^2$ as a function of the wind speed, which are available in the literature, are different, two model functions were considered. Firstly, according to the Cox-Munk formula [18], $\sigma=9.4°$ for the wind speed $U_{10}=3.4$ m/s at a height of 10 m.

However, the use of the total slope dispersion in this case is a too high estimate. A more accurate estimate can be obtained by using slope dispersion $\sigma^2_{\Delta f}$ in a frequency band $\Delta f$, in which the wave phases can be matched, for example, in the one-third octave band. After substituting dispersion relation (12) in the Toba spectrum (17) (which is in good agreement with the experimental data, as is seen in Fig. 3), a frequency-independent expression can be derived. According to the latter, $\sigma_{\Delta f}= 0.7, 1$, and $2°$, respectively, for the wind parameters $u_*=0.1$ m/s and the frequency bands $\Delta f = 0.2, 0.4$, and $0.6$ Hz (widths of the one-third octave band centered at 0.8, 1.6, and 2.5 Hz).

Figure B1 shows the results of calculations of the quadratic-to-linear harmonic amplitude ratio for the indicated values of the standard slope deviations $\sigma$ and at different viewing angles $\alpha_o$. The plots show that the optimal sector of viewing angles is $\alpha_o=45$ and $75°$. For the specified angular sector, the level of nonlinear harmonics in the one-third octave band is lower than 20 dB for the conditions of the experiment (at least, if the mechanism of the formation of surface brightness is due to changes in the reflection coefficient described by the Fresnel formula).

**Acknowledgments**

The authors would like to thank Dr. P.I. Korotin, Dr. A.V. Slunyaev, A.V. Ermoshkin, Dr. E.L. Borodina, Dr. A.G. Luchinin and foreign colleagues Dr. R.C. Spindel, Dr. G. Deane, and Prof. S. Cozzini for helpful discussions and valuable advices. The authors are grateful to A. Krayev for help with the English text. Also, the authors are grateful to the members of the Physical Acoustics Department of the Institute of Applied Physics RAS for help with the experiments.

**Figure captions**

1. Arrangement scheme of the angles with specular reflection of light from a patch of a rough surface $\Delta s$ when the sloping angle the surface patch lies in the longitudinal (Fig. 1a) and transverse (Fig. 1b) directions; $\alpha_o$ is the angle of the camera lens relative to the vertical.

2. (a) Video image of the sea surface: the red trapeze indicates the processing area, which, after the perspective correction, is projected into a rectangle, the white arrows point to the spar buoys; (b) the scheme of spar buoys (*1*- marks to determine the scale, *2* - water surface, *3* - buoyancy, and *4* - rope to anchor); (c) the zoomed image of a spar buoy obtained with the second camcoder; (d) frame (c) after the video signal processing on the PC video (the algorithm identified the above-water part of the spar buoy and determined the current value of the water level $h(t)$ at the spar buoy location).

3. The wind-wave PSD $G^2(\Omega)$ measured by the spar buoy and recovered by the brightness field versus the theoretical function corresponding to the Toba and Pierson-Moskowitz model spectra. Figures 3 (a)-3(c) show the spectra calculated from the experimental data I-III, respectively. The values of $u_*$, which were used to plot the Toba spectrum, and the respective values of $U_{10}$ (m/s): (a) $u_*=0.10$, $U_{10}=3.4$; (b) $u_*=0.08$, $U_{10}=2.8$, (c) $u_*=0.32$ and $U_{10}=8.9$. The frequency range $\Omega_1..\Omega_2$, for which the parameters were optimized (Hz): (a) 0.4..2; (b) 0.85..3; (c) 0.5..2. The values of $\gamma$ for the spectra recovered by the brightness field: (a) 1/6; (b) ¼; (c) ¼ for solid lines and $\gamma = ½$ for dotted lines.

4. The cross sections of 3D PSD of wind waves (surface waves) $G^2(\mathbf{K},\Omega)$ (in dB rel. 1m$^4$/Hz), white (black) circles (ellipses) limit the values of $|\mathbf{K}|$ satisfying the dispersion relation within the limits of a resolution of the spectrum analysis channel. The equal-level lines are plotted with a 5-dB step.

5. Average in the angular sector -60°..-30° spectral characteristics obtained from the video record I: (a) the PSD of surface brightness $I^2(|\mathbf{K}|,\Omega)$ in dB rel. not normalized units and (b) the PSD of surface wind waves $G^2(|\mathbf{K}|,\Omega)$ in dB rel. 1 m$^4$/Hz. The angles are measured from the axis $K_x$ counterclockwise (+90° is the viewing direction). L is the curve of the dispersion relation with the sleeves, corresponding to the spectral analysis resolution, Q is the curve of the second harmonic, A and B are the areas of the combination harmonics, which correspond to the difference (A) and the sum (B), and S is the boundary of A.

6. The PSD of wind waves (in dB) as a function of the phase velocity at a fixed wave vector, normalized to the maximum, averaged in the sectors -45..-15° (solid lines) or 30..60° (dashed lines). $K$ is given in 1/m, Record II.

7. Frequency dependence of the correlation coefficient modulus of the surface oscillations measured by a spar buoy and calculated by the brightness field, using (a) different types of calibration functions and (b) the calibration function ~i/(**Ks**) for fixed and frequency-dependent directions of the vector **s**, Record II. The analysis window length is 10 s and the averaging is over 16 windows. Vertical lines show the range where the level of the wind-wave PSD is not lower than 10 dB from the maximum.

8. The correlation coefficient modulus of surface oscillations measured by a spar buoy and calculated by the brightness field. Direction of vector **s**, which is included in the calibration function, is being scanned. The frequencies are 1.0, 1.3, 1.4, and 2.1 Hz. The calibration function is chosen as (a) ~i·sign(**K·s**)/$K$ and (b) ~i/(**K·s**). The zero degree is orthogonal to the camera axis.

B1. Estimation of the quadratic-to-linear harmonic amplitude ratio in the image spectrum of a wave with slope dispersion $\sigma^2$ in the case of the slope-brightness conversion according to the Fresnel formula, plotted in dB scale (20 log of amplitude ratio). The curves terminate before reaching the points 0 and 90°, where the incident ray is outside the 0-90°sector.

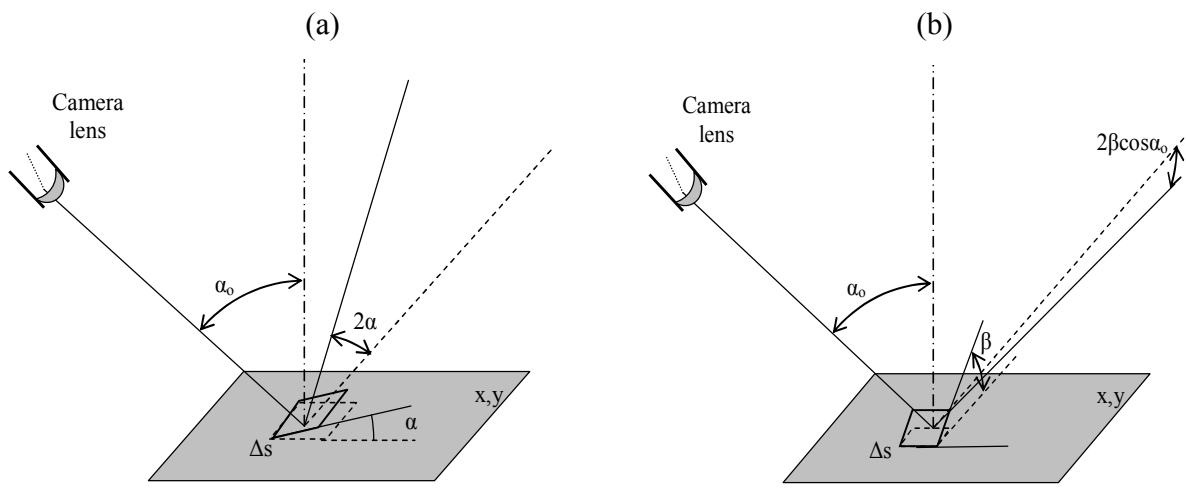

Fig. 1.

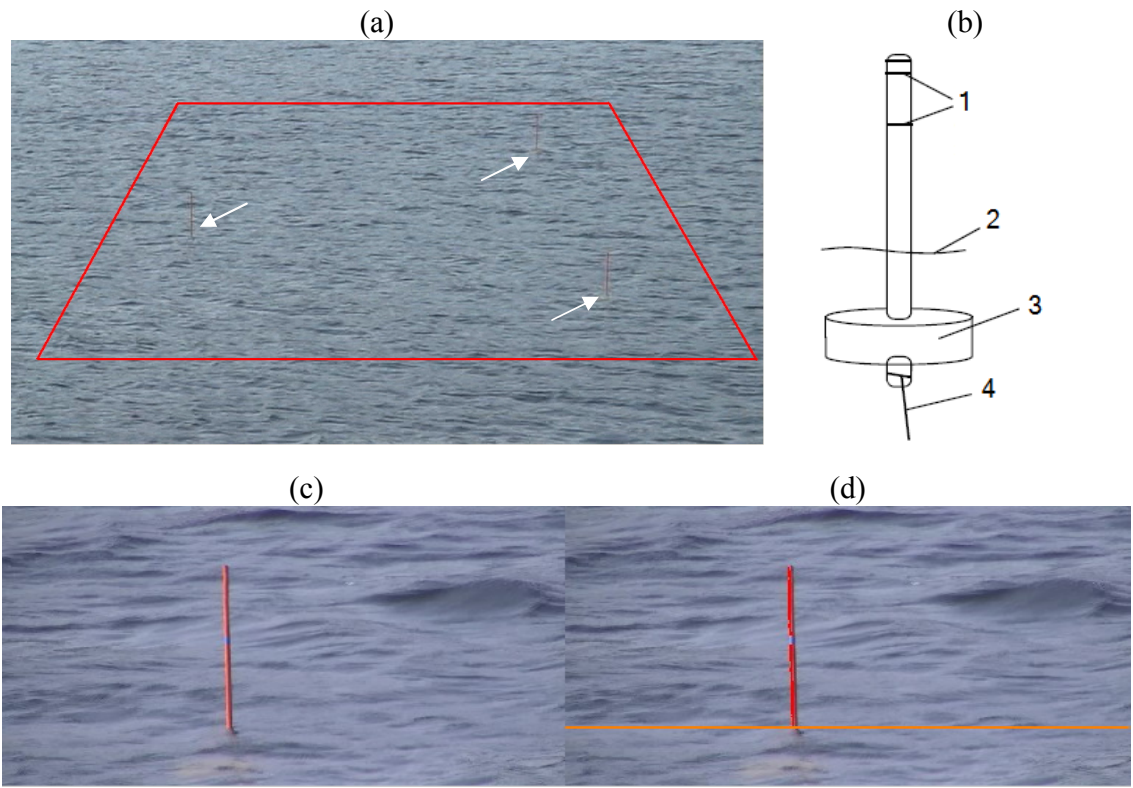

Fig. 2.

(Can be converted into gray scale in a standard way).

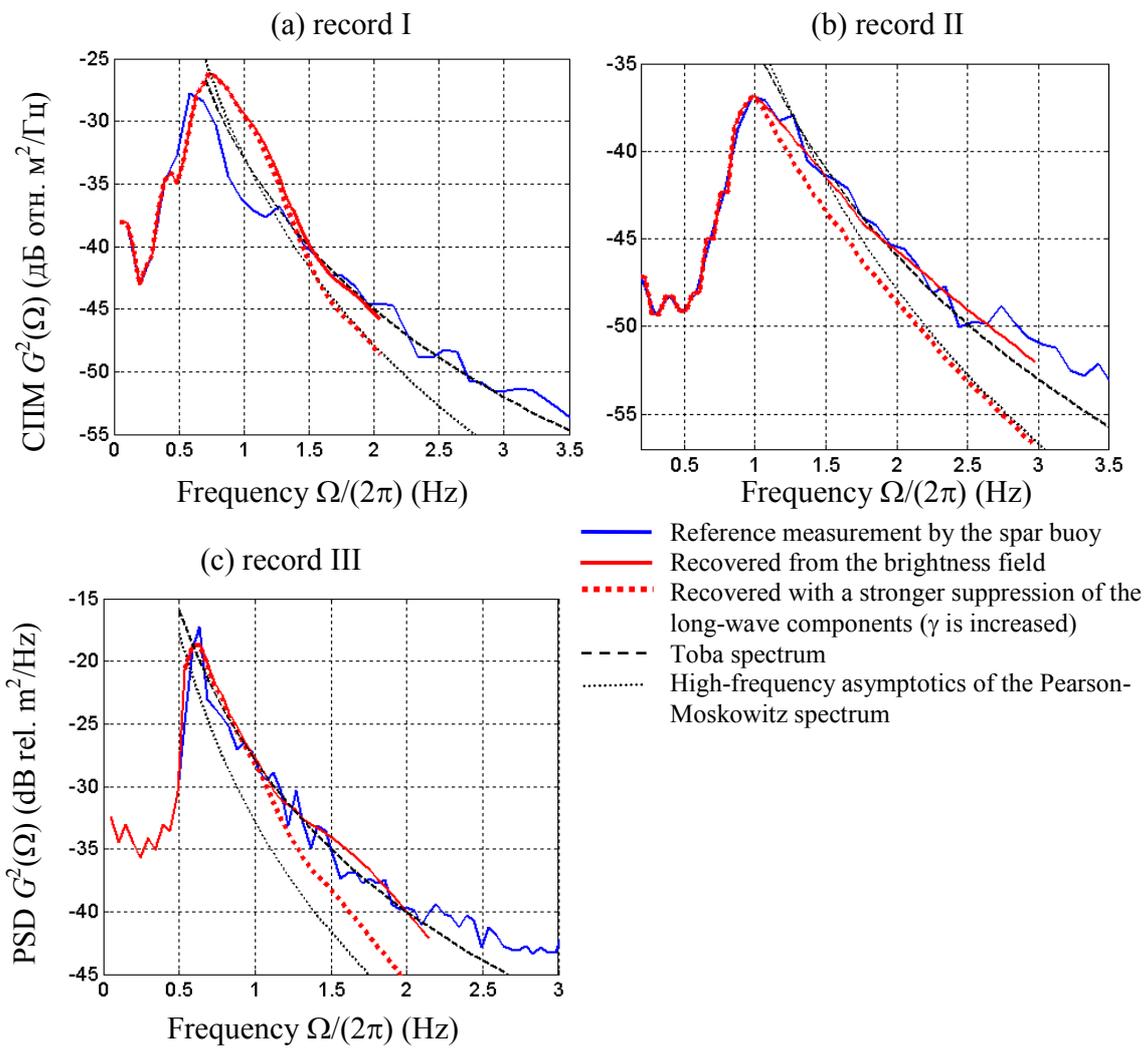

Fig. 3.

(Grayscale version is provided below)

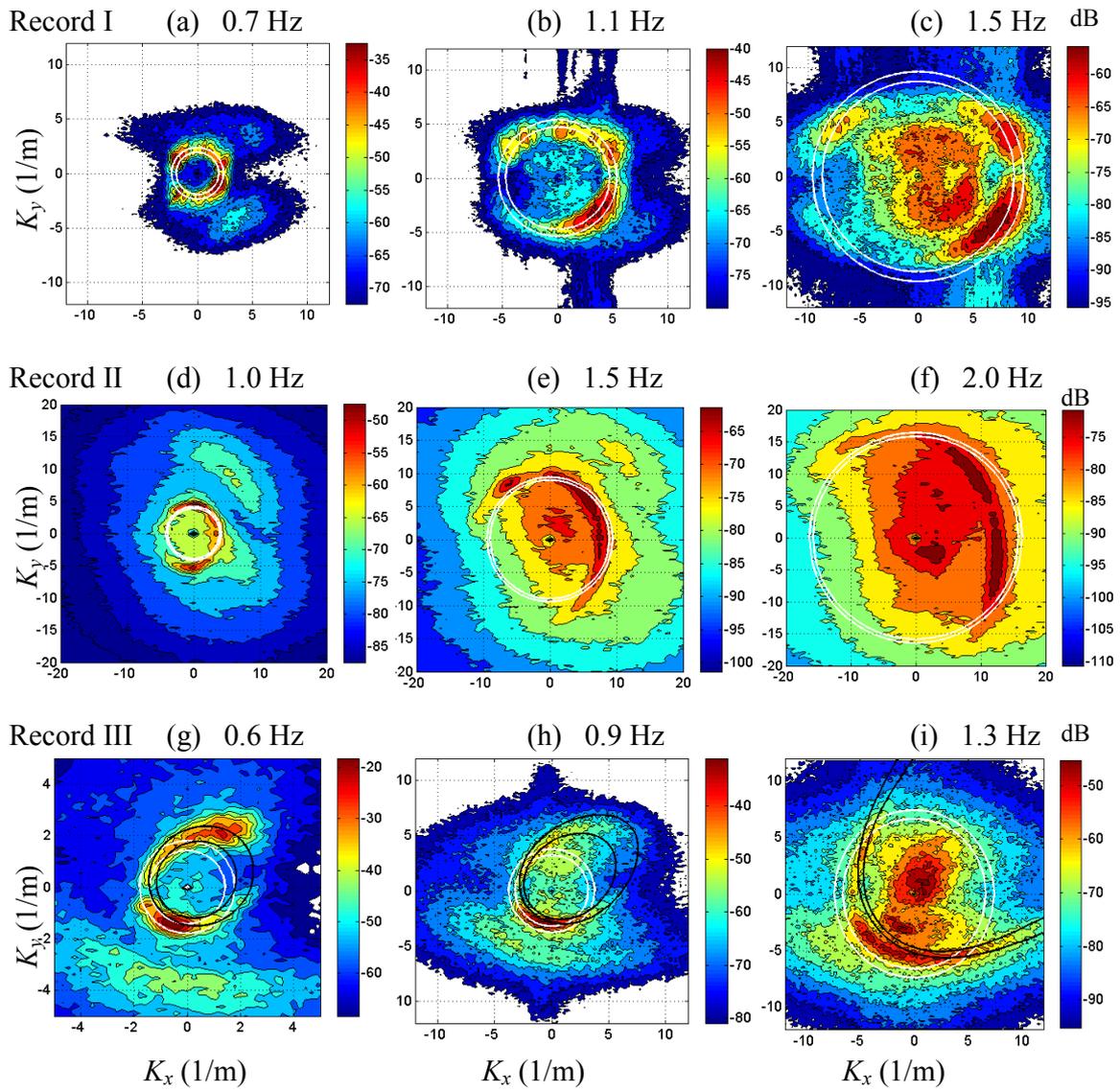

Fig. 4.

(Grayscale version is provided below)

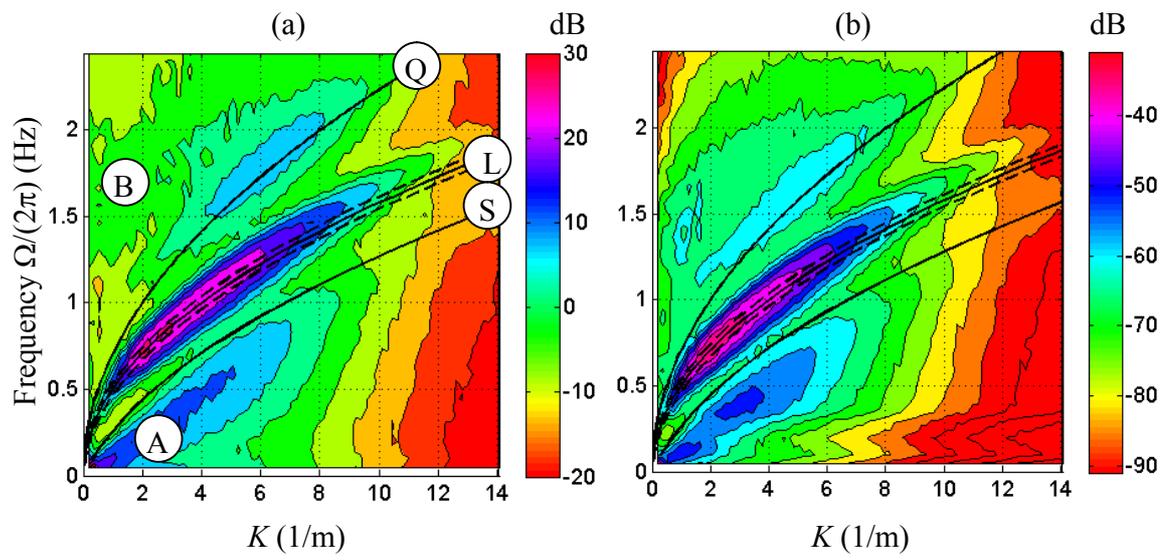

Fig. 5.

(Grayscale version is provided below)

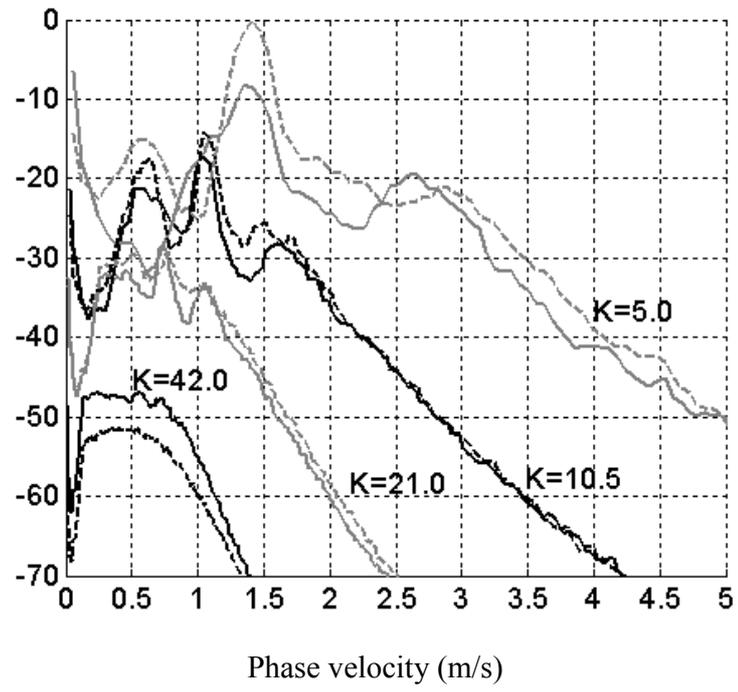

Phase velocity (m/s)

Fig. 6.

(a) 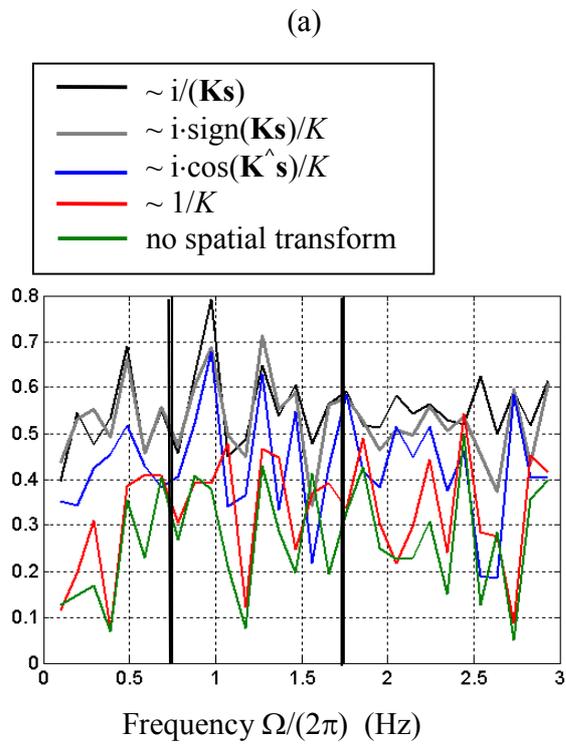

(b) 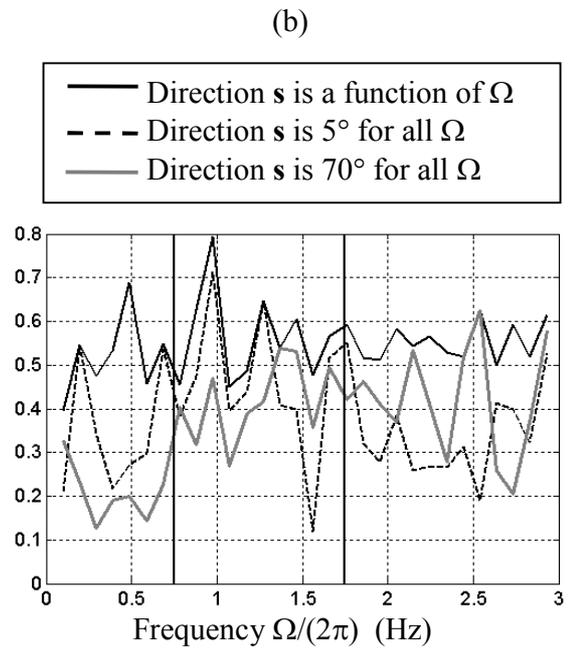

Fig. 7.

(Grayscale version is provided below)

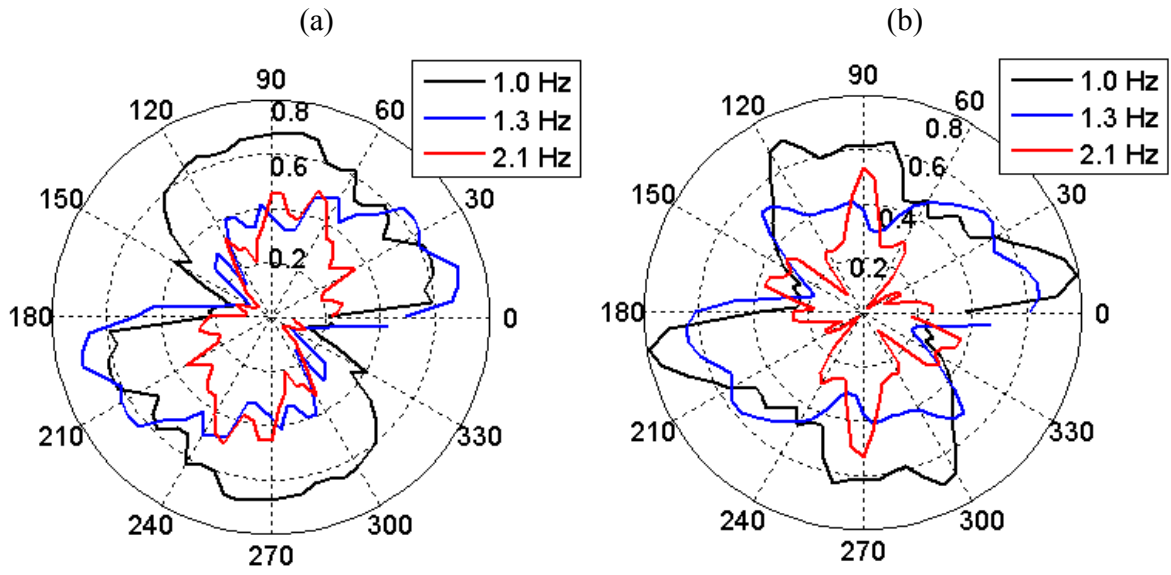

Fig. 8

(Grayscale version is provided below)

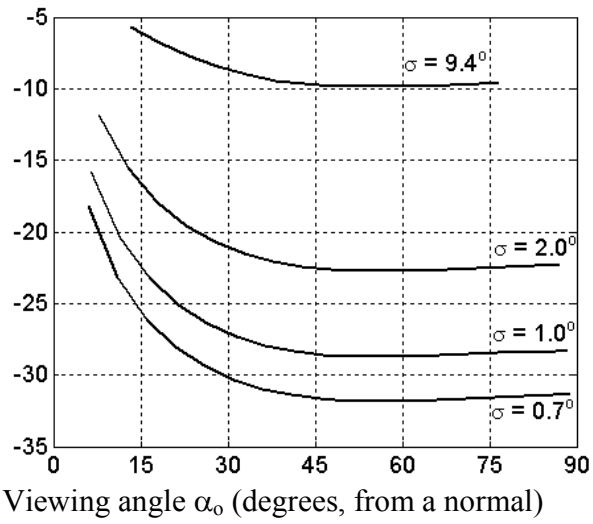
Viewing angle $\alpha_o$ (degrees, from a normal)

Fig. B1

Table 1.

| No. | I | II | III |
|---|---|---|---|
| Location, date | Baltic, 21.08.2006 | Ladoga, 19.07.2010 | Baltic, 19.08.2006 |
| Significant wave height[15], $H_S$ (m) | 0.1 | 0.06 | 0.2 |
| Frequency of the spectral peak $\Omega_P/(2\pi)$ (Hz) | 0.7 | 1.0 | 0.63 |
| Size of the processing area, $L_x$, $L_y$ (meters and number of peak wavelengths) | 35x57 (9x15$\lambda_p$) | 8x23 (5x14$\lambda_p$) | 30x40 (7x10 $\lambda_p$) |
| Analysis window duration $T$ (s) | 20 | 40 | 20 |
| Total record duration (min) | 2 | 15 | 2 |
| Minimum measured wavelength $\lambda_{min}$(m) | 0.4 | 0.2 | 0.4 |
| Height of the camera location above the sea level (m) | 18 | 5.8 | 18 |
| Shooting angle relative to the normal (degrees) | 78 | 77 | 81 |

---

[15] The classical definition is the standard deviation of the surface level, times 4.